\documentstyle[12pt,psfig,a4]{article}
\begin{document}
\newcommand{\mv}{$m_V$}
\newcommand{\beq}{\begin{equation}}
\newcommand{\eeq}{\end{equation}}
\newcommand{\beqn}{\begin{eqnarray}}
\newcommand{\eeqn}{\end{eqnarray}}
\newcommand{\adelt}{{\bf\Delta_t}}\newcommand{\akt}{{\bf k_t}}
\newcommand{\cpt}{c_{\phi-\theta}}
\newcommand{\bl}{\beta}\newcommand{\cp}{c_\phi}
\newcommand{\alt}{{\bf
l_t}}\newcommand{\ct}{c_\theta}\newcommand{\st}{s_\theta}
\newcommand{\mqt}{m_q^2}

\noindent\hbox to\hsize{July 1998, revised December 1998\hfill
ULG-PNT-98-2-JRC}\\
\vskip 1.6in
\begin{center}
{\bf FERMI MOTION AND QUARK OFF-SHELLNESS\\ IN ELASTIC VECTOR MESON PRODUCTION
}\\
\vskip 0.3in
 Isabelle Royen\footnote{iroyen@ulg.ac.be}
and Jean-Ren\'e Cudell\footnote{JR.Cudell@ulg.ac.be}
\\{\small
Inst. de Physique, U. de Li\`ege,
B\^at. B-5, Sart Tilman, B4000 Li\`ege, Belgium\\
}

\vskip 0.3in
{\bf Abstract}
\end{center}
\begin{quote}

We study ways of implementing Fermi momentum in elastic \break
vector-meson production, and find that the usual on-shell assumption
of quark models and quark wave functions cannot reproduce the ratio
$\sigma_L/\sigma_T$.
We propose a new approach which allows the quarks to be off-shell,
and which naturally reproduces the data. As a consequence, we prove that
the asymptotic form of the transverse cross section is different
from $\sigma_T\sim 1/Q^8$.
In this new model,
we show that the mass, $t$  and $Q^2$ dependence of the cross sections are also
reproduced. We also make predictions concerning the production of excited
states such as the $\rho'$ and the $\psi'$.

\end{quote}

\newpage
\section{Introduction}
The data from HERA
have now reached a remarkable level of accuracy, and in particular they
provide us with excellent information on vector meson elastic production
 \cite{Zeusdat,H1dat}.
We have precise measurements of the photon $Q^2$, $t$ and $w^2$ dependences
of the cross section, for different mesons, as well as information
on its transverse and longitudinal components, $Q^2$ being the absolute value
of the photon off-shellness, $t$ the square of the momentum transferred
to the proton and $w^2$ the square of the center-of-mass energy of the
$\gamma^{(*)} p$ process.

Theoretically, this is related to a measurement of the off-diagonal
deep-in\-elas\-tic tensor $W^{\mu\nu}(x,Q^2,m_V^2,t)$, for $t$ non zero. One
may
hope that the presence of two new scales - the vector meson mass $m_V$ and the
momentum transfer $t$ would enable one to concentrate on regions where the
perturbative calculation is reliable, and observe how one can
extend it to smaller scales.
Theoretically, these studies have recently been set on a firm ground as
Collins,
Frankfurt and Strikman \cite{Collins} have shown that a factorisation theorem
exists for exclusive vector meson production. This theorem asserts that, to
leading power of $Q^2$ and to all logarithms, the cross section is the
convolution
in longitudinal momentum of a hard scattering amplitude, an off-diagonal
structure function and a meson wave function. Furthermore, the theorem can be
proven irrespectively of the meson mass. Hence at high $Q^2$, even light meson
production can be reliably calculated.

The $w^2$ dependence of the cross section then results from the $x$, $t/w^2$
and
$m_V^2/w^2$
dependence of these new distribution functions. It is to be noted that these
are new objects which are different from the diagonal structure functions.
Also, the theorem naturally holds for longitudinal vector meson
production, which is leading in $Q^2$, and for which one can show that the
off-shell distributions are related to the usual gluon distribution. The
transverse
cross section $\sigma_T$ is {\it a priori} down by a factor $Q^2$. However, the
factorisation theorem envisions the possibility of a different behaviour, but
the price to pay, in the case of massless quarks, is to allow the leading
behaviour of $\sigma_T$ to come totally from the non-perturbative region, for
quark off-shellnesses of order $m_V^4/Q^2$.
In this paper,  we shall go beyond this analysis by including massive
quarks in a  model describing the transition
from a photon to a QCD bound state. One indeed needs to disentangle
the effect of this transition on
the transverse cross section, before one can extract the off-diagonal
distributions accurately.

One expects (and we confirm) that the transition $\gamma^* \rightarrow V$
will not modify the $w^2$ dependence significantly at high energy. The
$w^2$ dependence is surely one of the outstanding problems of QCD, and
our theoretical understanding of it seems to have made a big step recently
\cite{NLL}, although it is difficult to assert yet in which direction. We
shall make the simplifying assumption that it enters as a constant factor
which does not depend on any variable but $t$. We shall come back later to
the significance of this assumption.

We could of course, as often done
nowadays, assume that this factor is related to $W(t=0)$ and is proportional
to $[xg(x)]^2$. However, we want first to point out that this correspondence
exists only for longitudinal cross sections, and is only approximate.
Indeed, as we shall explain below, the imbalance
between initial and final states in vector-meson elastic production brings one
to a different kinematic
domain from DIS. This has important consequences for quantities related to
the upper loop of a ladder, especially in the transverse case.

Hence we shall simply calculate up to a factor, and we shall not claim that we
can reliably predict it. On the other hand, the $Q^2$ and $m_V$ dependences
mainly (up to logarithms) come from the upper loop. As we shall see, the data
is not precise enough (or $Q^2$ not large enough) to require the inclusion of
evolution effects. As for the $t$ dependence, it comes from the proton form
factor, the $\gamma^* V$ loop, and $W$. We shall show here that simple
assumptions enable one to reproduce the data at small $t$, and that
high-$t$ points do not seem to require either a pomeron slope
or a BFKL enhancement.

Our starting point is a very simple model that we published recently
\cite{CR1}. It is
based on lowest-order perturbative QCD and
on a very na\"\i ve approximation to the meson wave function \cite{Horgan},
where in its rest frame the meson is simply described by two quarks at rest.
Such an approximation should be valid provided that the meson motion is
negligible, {\it i.e.} at very high $Q^2$.
Despite its simplicity, this model was surprisingly successful and reproduced
 most of
the features of the data: the mass dependence of the cross section came out
naturally, as well as the $Q^2$ dependence. The model also predicted correctly
s-channel helicity conservation, and even worked in photoproduction.
It provided a test of two ideas: first of all,
elastic production is mainly due to two gluons interacting with the meson,
secondly, as the mesons were supposed to be made of their constituent quarks,
the main production mechanism involves only the lowest Fock state.

Despite its many successes, this model
failed, as most other, in the fact that the ratio $\sigma_L/\sigma_T$ was
predicted to be a factor 8 higher (at $Q^2=20$ GeV$^2$) than observed
experimentally. As $\sigma_L$ is much larger than $\sigma_T$, the high-$Q^2$
total cross sections $\sigma_L+\sigma_T$
were unaffected by this failure, but it is a strong indication that the model
is far from complete. We want to emphasize that so far no model manages to
reproduce the plateau observed experimentally, and that the best one has been
able to do is to get a slower linear rise.

We shall show here that a re-consideration of the role of Fermi momentum
enables one to preserve all the previous successful results while reproducing
the ratio $\sigma_L/\sigma_T$. We keep the two fundamental assumptions of
our previous paper, that the process is dominated by the coupling of two
gluons to the lowest Fock state. In Section 2, we introduce the general
formalism which we shall use for the description of the quarks contained in the
meson.
In sections 3 and 4, we give the general form of the amplitude. In section 5,
we
show that the amplitude is in general infrared finite, and in section 6 we
give our results, and discuss the off-shell contribution to the amplitude.
We then conclude, and pose several questions concerning the validity of
the wave function formalism, and of the connection between diagonal and
off-diagonal DIS tensors.

\section{Vertices vs. wave functions}
We shall assume here a form of factorisation, in that we shall not consider
soft exchanges between the proton and the meson, but only concentrate on
a hard scattering, a meson vertex function, a singlet exchange and a proton
form factor. We do keep the exact kinematics, including transverse motion
and quark masses, in the amplitude describing the transition
$\gamma^*\rightarrow V$ as we want to assess the contribution of near-shell
partons.

We need a model for the off-diagonal structure function, and we adopt the
simplest one: at high energy, the process is dominated by pomeron exchange,
which
we model by two gluons times a Regge factor. The model of the off diagonal
structure function will be obtained by convoluting the exchange with a proton
form factor, which partially kills the infrared divergences.

The hard process generating the meson, and the $q\bar q\rightarrow V$ amplitude
are treated together through the introduction of a meson vertex function.
We shall make the assumption that the gluons couple to the constituent
quarks of the meson, and assume that the direct coupling of gluons to
the $V$ vertex is negligible. In general, this is not the case, as
the vertex must result from a non-perturbative resummation of diagrams
involving quarks and gluons, to which perturbative gluons can in general couple
directly. In fact, in all generality, such diagrams are needed to obtain a
gauge invariant amplitude \cite{Diehl,LandHe}. However, as we shall see, these
terms do not contribute in the high-energy limit.

Hence we are left with the description of the $\bar q q V$ vertex, and
the gluons will couple to the quarks emerging from this vertex.
In general, this vertex can depend on
the 4-momenta respectively of the vector meson, of the quark, and of the
antiquark,
$V\equiv 2v$, $v+l$ and $v-l$, as well as on $\gamma_\mu$.
We know experimentally that vector-meson electroproduction conserves
s-channel helicity, hence the only possible tensor structure is $\gamma_\mu$.
We are thus left with a vertex:
\beq \Gamma_\mu=\Phi(l) \gamma_\mu\eeq
We shall assume that the vertex function $\Phi$ can depend only on the relative
4-momentum $l$, and that
the dependence on the meson mass can be restated as a dependence on a Fermi
momentum scale $p_F$.

This vertex function can be formally related \cite{LandHe}
to a light-cone wavefunction, but
we shall not make use of such a wavefunction in the meson case. Indeed,
we find that the analytic structure of the propagator which is
usually included in the definition of the
wavefunction plays a crucial role in the reproduction
of the ratio $\sigma_L/\sigma_T$. In the following, we shall make no further
assumption concerning the quarks, except that we can treat them within
perturbation theory. In particular, we shall not treat them
{\it a priori} as on-shell particles.

In the following, we shall assume that the function $\Phi$ can be approximated
for s states by a falling gaussian, multiplied by a normalisation constant $N$.
As we do not assume that quarks are real particles, we do not have the usual
wavefunction  normalisation
integral on $\Phi$. However, the same vertex controls the decay $V\rightarrow
e^+e^-$, and hence we shall be able to determine $N$ so that it correctly
reproduces the latter rate.

\section{Calculation of the amplitude}
\subsection{Kinematics}
The vectors of the problem are defined in Fig.~1. The photon has 4-momentum $q$
and polarisation $\epsilon$, with $q.q=-Q^2$. The vector meson has momentum
$V=q+\Delta$ and
polarisation $e$, with  $t=\Delta^2$,  $V^2=m_V^2$, hence
$\Delta.q=(m_V^2-t+Q^2)/2$.
The quarks composing the meson
are written as $v+l$ and $-v+l$ with $v=V/2$.
\begin{figure}[here]
\centerline{
\hbox{
\psfig{file=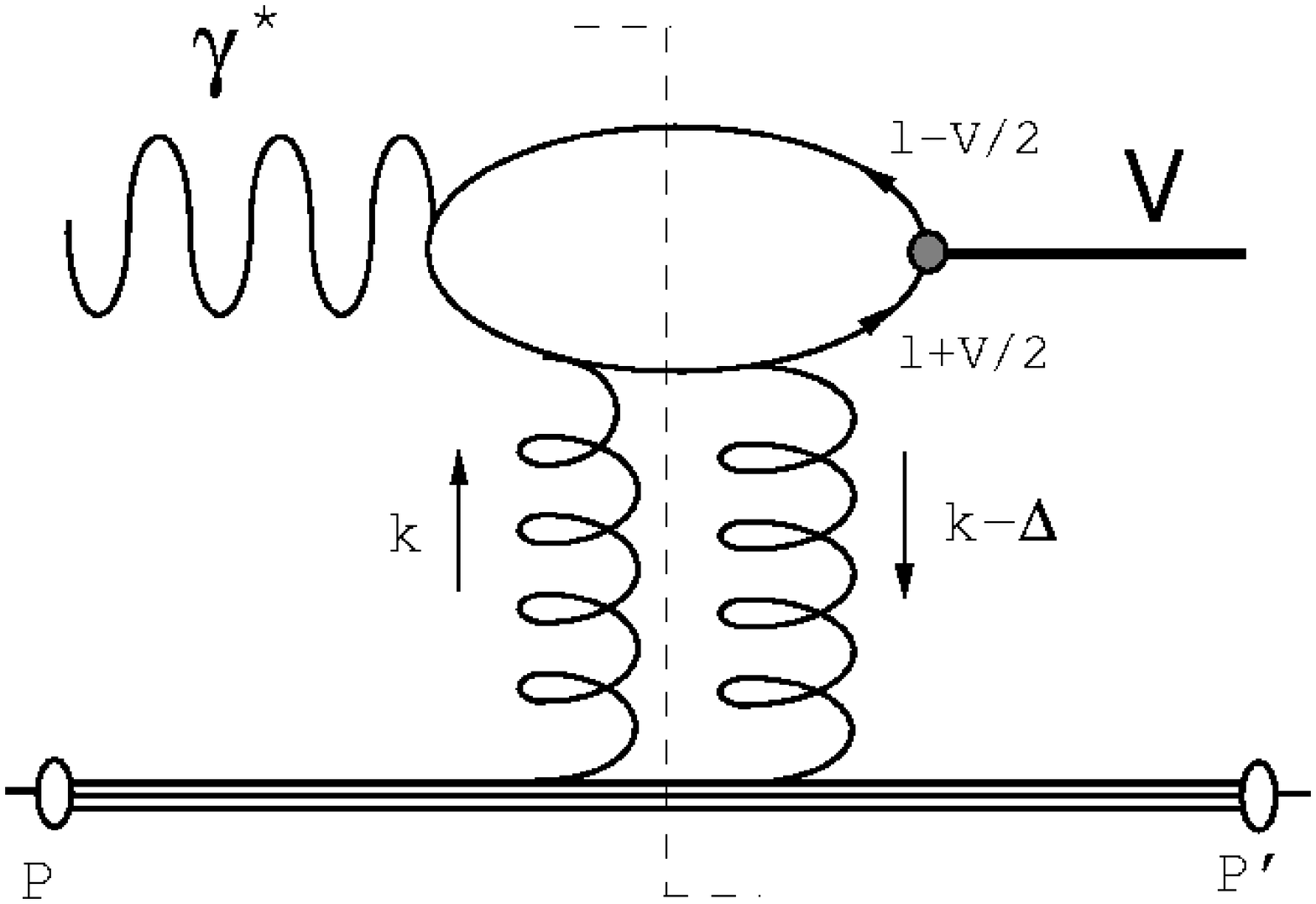,bbllx=2.5cm,bblly=13cm,bburx=20cm,bbury=23cm,width=5cm}
\hglue 1cm
\psfig{file=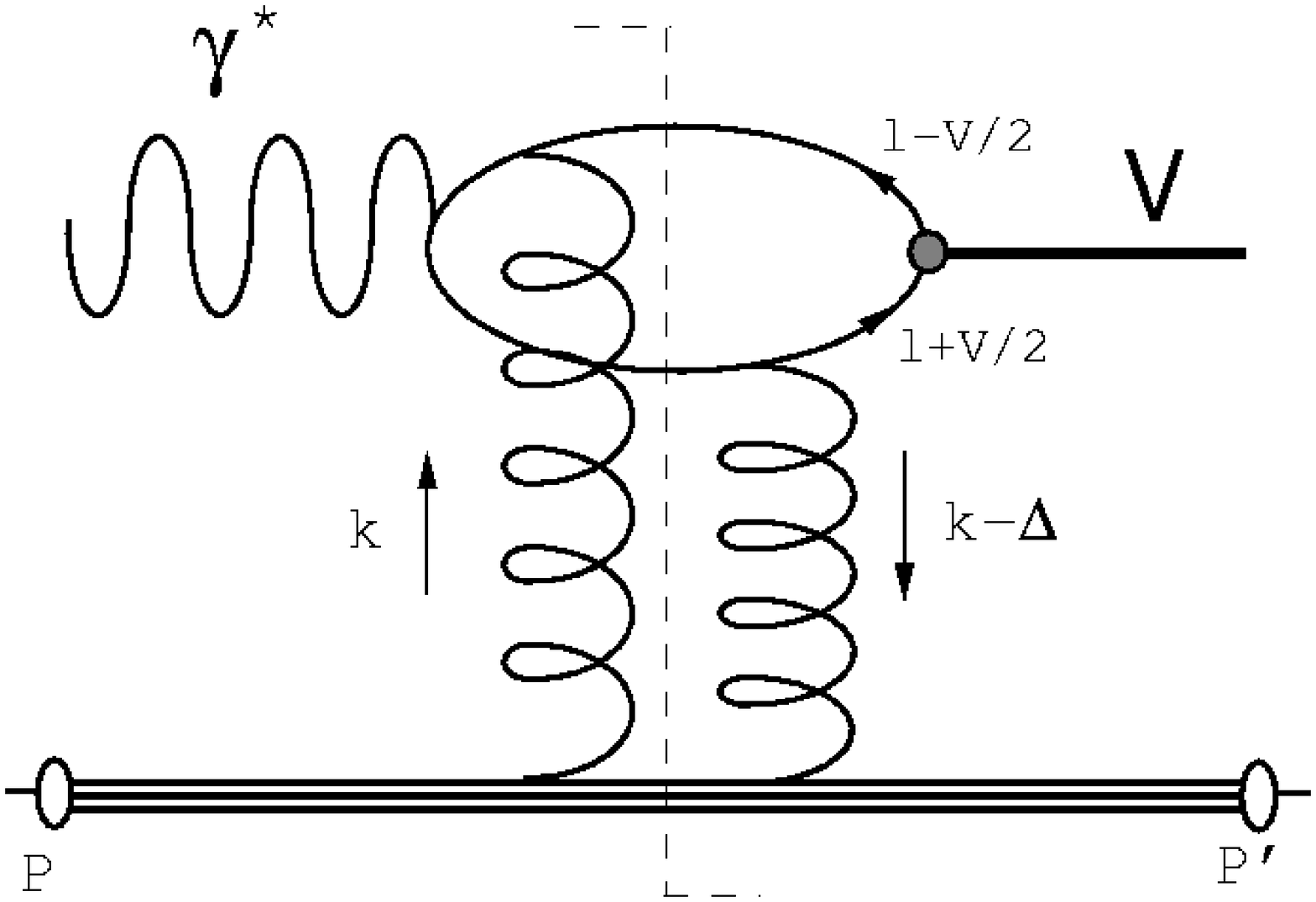,bbllx=2.5cm,bblly=13cm,bburx=20cm,bbury=23cm,width=5cm}
}}
\vskip 1in
\begin{quote}
{\small Figure~1:  The two diagrams accounting for the transition
$\gamma^*p\rightarrow Vp$.
The dashed line represents the cut which puts the intermediate state on-shell.
}\end{quote}
\end{figure}

As we do not want to look in detail at the diffracted proton, we follow the
usual treatment \cite{GS}:  the momentum of the quark is $p$, and we assume
that
we can neglect its mass and put it on-shell $p.p=0$.
The fact that the proton quark remains on-shell after the scattering implies
 $p.\Delta=t/2$.
We shall be working in the high-$w^2$ limit, and we write
$p.q\approx(p+q)^2/2\equiv \hat w^2/2$. We define $\hat w\approx w/3$ at the
quark level, but this will not make any
difference, as the final answer, in the large $\hat w$ limit, will be
independent of $\hat w$.
\vfill\break
\subsection{The photon-meson bubble}
\renewcommand\thefootnote{\dagger}
Assuming that we can use perturbative quarks\footnote{One could easily extend
this formalism and include non-perturbative effects present in the quark
propagator, by replacing the quark mass $m_q^2$ by the parameters that enters
in
the Hilbert transform of the propagator, and by integrating over
the Hilbert density afterwards. Such a complication is not needed to
describe the data.} in the upper bubbles, or equivalently that the
non-perturbative effects can be reabsorbed into the vertex function, the upper
bubbles of the graphs are described by the following traces:
\beqn
T_1^{\alpha\beta}&=&Tr\{\Phi\left(l\right)\ \gamma.e [\gamma.(v+l)+m_q]
\gamma^\beta [\gamma.(q-v+l+k)
+m_q]
\gamma^\alpha\nonumber
 \\&\times&[\gamma.(q-v+l)+m_q] \gamma.\epsilon [\gamma.(-v+l)+m_q]\}\\
T_2^{\alpha\beta}&=&Tr\{\Phi\left(l\right)\gamma.e [\gamma.(v-l)+m_q]
\gamma^\alpha
[\gamma.(v-k-l)+m_q]
\gamma.\epsilon \nonumber\\
&\times& [\gamma.(v-q-k-l)+m_q] \gamma^\beta [\gamma.(-v-l)+m_q]\}
\eeqn
One of the quark lines connecting to the photon
is off-shell and has different
expressions in each diagram. Its propagator in the first graph is
\newcommand{\dm}{m_V^2/4-m_q^2}
\newcommand{\sm}{m_V^2/4+m_q^2}
\beq P_1= (q-v+l).(q-v+l)-m_q^2=(l.l + 2 l.q - 2 l.v - 2 q.v +{m_V^2 \over
4}-m_q^2 -
  Q^2)\eeq whereas in the second graph it reads:
\beqn P_2&=& (-v+l+k).(-v+l+k)-m_q^2\nonumber\\ &=&
 (k.k + 2 k.l - 2 k.v +  l.l - 2 l.v +{m_V^2 \over 4}-m_q^2)\eeqn

As we expect the amplitude to be $w$-independent,
we shall be calculating the discontinuity of the amplitude, and the two
intermediate quark propagators are put on-shell, which gives us the relations:
\begin{eqnarray}
k.q &=& \Delta.k + \Delta.q - k.k - 2 k.l - 2 l.q \\
l.q &=& - \Delta.l +  l.l +\dm
\end{eqnarray}
One of the quarks constituting the meson is in general not cut,
hence we have another propagator
\begin{equation}
P_5=(v+l)^2-m_q^2= 2( l.l +\dm)
\end{equation}

The sum of the two cut diagrams will then give:
\beq
T^{\alpha\beta}=[{T_1^{\alpha\beta}\over P_1}+{T_2^{\alpha\beta}\over
P_2}]{(2\pi)^2 \delta(P_3) \delta(P_4)\over P_5}
\eeq

\subsection{Current conservation}
One of the consequences of gauge invariance is that the contraction of
the upper bubble with the momenta of external lines is zero.
In our case, we obtain indeed
$k_\alpha T^{\alpha\beta}=0$, and $T^{\alpha\beta} (\epsilon\rightarrow q)=0$.
However, we do not obtain exact current conservation for the off-shell quark
emerging from the meson vertex:
$(k-\Delta)_\beta T^{\alpha\beta}\sim l.v\neq 0$.
In the photon case, one can recover explicit current conservation at
the $\beta$ vertex by crossing both sides of the diagram \cite{CDL}. This can
be
accomplished by the substitution $T\rightarrow {1\over 2}[T(l)
+T(l\rightarrow \Delta-k-l)]$. In our case, however, because of the presence of
the vertex function, such substitution is not possible, and the current
conservation that could be checked explicitly at the $\beta$ vertex in the
photon case is in general absent.

We can however impose that current
conservation be satisfied, which amounts to putting the other quark making up
the vector meson on-shell.
The amplitude then splits into 2 physical processes:
$\gamma gg\rightarrow \bar q q$ and $\bar q q\rightarrow V$. This of course
leads to current conservation and ensures gauge invariance. This is the
approach usually followed in a wave function formalism. It has the drawback of
relying heavily on the existence of a mass shell for the quarks, and as we
shall see does not lead to observed results.

However, we can
pursue another route, by keeping the gauge dependence of the propagators,
and checking explicitly that it cancels out in the large $w^2$ limit. As we
shall see, the leading terms come from terms in which the momenta of
the upper bubble get contracted with momenta of the lower quark lines. This
means that terms proportional to $g_{\mu\nu}$ in the propagators are enhanced
by a factor $w^2$ with respect to terms containing $k_\mu$ or $k_\nu$, as the
gluon
momentum $k$ will turn out to be transverse. Hence the large-$w^2$ limit is
gauge invariant. We shall show explicitly that the infrared cancellation which
is expected from gauge invariance occurs explicitly, albeit after integration
over
the photon-meson bubble.

\subsection{The full amplitude}

In order to obtain the full amplitude, we need to add the contribution of the
proton lines, and the gluon propagators.
We represent the
proton by a constituent model, introducing the
two following forms factors \cite{CR1,GS}:

\begin{equation}
{\cal E}_1(t=\Delta^2)\approx \frac{(3.53-2.79t)}{(3.53-t)(1-t/0.71)^2} \\
\end{equation}
when both gluons hit the same quark line (this is a fit to the measured
Dirac elastic form factor), and
\begin{equation}
{\cal E}_2(k,k-\Delta)={\cal E}_1(k^2+(k-\Delta)^2-c\ k\cdot (k-\Delta))
\end{equation}
if the gluons hit different quark lines, with $c\approx 1$ \cite{CH}.\\
The prescription to go from the quark-level process to the proton-level process
is then to multiply the amplitude by the form factor ${\cal F}(k,\Delta)=
3 ({\cal E}_1-{\cal E}_2)$.
The leading contribution of the lower quark line being
$4 p^\alpha p^\beta$, we obtain the following expression for the
amplitude:
\begin{eqnarray}
{\cal A}&=& {2 \over 3}(4\pi \alpha_S)^2\ g_{elm} e_Q\nonumber\\
&\times&\int\ \frac{d^4l}{(2\pi)^4} \int\frac{d^4k}{(2\pi)^4}
{\cal F}(k,\Delta)\ {1 \over \sqrt{3}}\Phi\left(l\right)
\nonumber\\
&\times& \frac{[4 (p_\alpha p_\beta) (g^{\alpha\tau}+\lambda {k^\alpha
k^\tau\over k.k} )
(g^{\beta\mu}+\lambda {(k-\Delta)^\beta (k-\Delta)^\mu\over
(k-\Delta).(k-\Delta)})T_{\tau\mu}]}{k^2(k-\Delta)^2}\
\eeqn
where $\lambda$ is the gauge parameter and
$e_Q g_{elm}=e_Q \sqrt{4\pi \alpha_{elm}}$  the electromagnetic coupling
of the different vector mesons: $e_Q=\frac{1}{\sqrt{2}}$ for the $\rho$,
$-1/3$ for the $\phi$ and $2/3$ for the $J/\psi$.

\section{The high-energy limit}
To take the large-$w^2$ limit, we shall use Sudakov variables,
rewrite the vectors of the problem in terms of $p$ and $q$, and define
the transverse direction, noted with a ``t" subscript, as being orthogonal to
both $p$ and $q$.

We first write the expression for the polarisation vectors.
The
transverse photon polarisation is simply a unit vector in the transverse plane,
whereas the transverse polarisation of the vector is slightly more complicated,
as the transverse plane has been defined w.r.t. $q$.
Solving $e.e=-1$, $e.v=0$, $\epsilon.\epsilon=1$ and $\epsilon.q=0$, we obtain:
\begin{eqnarray}
\epsilon_L&=&{2Q \over w^2} p + {1\over Q} q\\
\epsilon_T&=&\epsilon_t\\
e_L&=&{\Delta_t\over m_V} + { - m_V^2 + Q^2 - t\over m_V w^2} p +
{t + w^2\over m_V w^2} q\\
e_T&=&e_t - {2\Delta_t.e_t\over w^2}p
\end{eqnarray}

Taking into account the on-shell conditions, we obtain
the following expansions for the measures and the 4-vectors:
\beqn
&&d^4\Delta \ \delta_+((p-\Delta)^2)\ \delta_+((q+\Delta)^2-m_V^2)={1\over
2w^2}\  d^2\Delta_t\nonumber\\
&&\Delta=\Delta_t + {m_V^2 + Q^2 - t\over w^2} p + {t\over w^2} q\eeqn

\beqn
&&d^4l\ \delta_+((v-l).(v-l)-m_q^2)={1\over 8(1 - \beta)}\ d\beta d^2l_t
\ \theta(1-\beta)
\nonumber\\
&&l={l_t\over 2} +{\beta\over 2} q\\
&&\ \ \ + {(2\Delta_t-l_t).l_t + \beta((\beta-1) Q^2 +  m_V^2 - t
)-m_V^2 + \mu_q^2 \over 2(\beta-1)\ w^2} p  \nonumber
\eeqn
with $\mu_q=2 m_q$.

\beqn
&&d^4 k\  \delta_+((q-v+l+k)^2-m_q^2)\ \delta_+((p-k)^2)={1 \over 4(\beta + 1)\
w^2}\ d^2 k_t\ \theta(1+\beta)\nonumber\\
&&k={k_t\over 2}+ {k_t^2\over 4w^2}q \\
&&+ { 2[(1- \beta^2)Q^2 - t + \mu_q^2 - l_t.(l_t-2\Delta_t)] - k_t.(-k_t + 2l_t
- 2\Delta_t)(1-\beta)
\over 2(1-\beta^2) w^2}p\nonumber
\eeqn

The propagators then become:
\beqn
P_5&=&2l^2 + {m_V^2 - \mu_q^2 \over 2}\nonumber\\
k^2&=& {k_t^2\over 4}\nonumber\\
(k-\Delta)^2&=& - \Delta_t.k_t + {k_t^2\over 4} + t\eeqn

In the following, we shall write the integrated amplitude as:
\begin{eqnarray}
{\cal A}_{(T,L)}&=& {2 \over 3}(4\pi \alpha_S)^2\ g_{elm} e_Q\nonumber\\
&\times&\int\ \frac{d^2k}{(2\pi)^2} \int\frac{d^4l}{(2\pi)^4}\nonumber\\
&\times&{\cal F}(k,\Delta)\ {1 \over \sqrt{3}}\Phi\left(l\right)
\ \bar A_{(T,L)} \eeqn

The leading terms in $w^2$ for the two amplitudes $\bar A_{(T,L)}$ then become
$\bar A_T= N_T/D$ and $\bar A_L= N_L/D$ with
\beqn D&=&
(2\beta \Delta_t.l_t -l_t.l_t  + \mu_q^2 + (\beta^2 - 1)m_V^2 -
t\beta^2)\nonumber \\
&\times& ((1-\beta^2)Q^2 + \mu_q^2 - t -(l_t+k_t-2\Delta_t).(l_t+k_t))\nonumber
\\
&\times& ((1-\beta^2)Q^2 +(2\Delta_t - l_t).l_t + \mu_q^2 - t)\nonumber \\
&\times& (4\Delta_t.k_t +k_t.k_t - 4t)k_t.k_t\\
N_L&=&
{4(1-\beta^2) \ Q \over m_V}
[m_V^2(\beta^2-1) + t\beta^2 - \mu_q^2 +l_t.(l_t - 2\beta\Delta_t)]\nonumber \\
&\times& k_t.(2l_t + k_t - 2\Delta_t)\\\nonumber\\
N_T&=&8 {\epsilon_t}_\mu {e_t}_\nu \nonumber\\
&\times& \{[-\Delta_t^\mu \Delta_t^\nu  (\beta + 1) \beta +\Delta_t^\mu l_t^\nu
\beta - l_t^\mu l_t^\nu (\beta + 1)+ l_t^\mu \Delta_t^\nu (\beta^2 + \beta +
1)] \nonumber\\
&& \ \ \ \ (k_t + 2 l_t - 2 \Delta_t).k_t (1-\beta)  \nonumber\\
&&+[ - \beta \Delta_t^\mu k_t^\nu -k_t^\mu l_t^\nu \beta^2 +k_t^\mu
\Delta_t^\nu \beta^3 +l_t^\mu k_t^\nu] \nonumber\\
&&\times[t - \mu_q^2 - (1-\beta^2) Q^2 + l_t.l_t - 2 \Delta_t.l_t]\nonumber\\
&&- g^{\mu\nu}[ - k_t.(k_t + l_t) (l_t.l_t -\mu_q^2) -(1 - \beta^2) k_t.l_t Q^2
\nonumber\\
&&\ \ \ +\Delta_t.\left((\mu_q^2  -l_t.l_t) (\beta-2) k_t + ((\beta + 1)
k_t.k_t + 2\beta k_t.l_t -2 \Delta_t.k_t) l_t\right) \nonumber\\
&&\ \ \ +t k_t.(\Delta_t \beta - k_t \beta - 2 l_t \beta + l_t) \nonumber\\
&&\ \ \ + (1-\beta^2) \beta \Delta_t.k_t Q^2]\}\label{tpart}
\eeqn

In order to proceed further, we choose two orthogonal polarisation vectors
in the transverse case, and
there are a priory four transverse amplitudes, as the polarisation vector can
each be taken parallel or orthogonal to a fixed direction. However, we find
that the non diagonal terms, in which the photon and the vector meson
have orthogonal polarisations, cancel in the angular integration.
Hence in general we have only two diagonal transverse amplitudes
to consider.

\section{Properties of the amplitudes and of the cross section}
To pursue analytically, we shall concentrate on the case $t=0$ and $\Delta =
(m_V^2 + Q^2)/w^2 p $. We shall
give numerical results at all $t$ further on. We can then perform
the angular integrals analytically.
In this case, the remaining two transverse amplitudes become
equal after angular integration.
Hence we are left with a transverse and longitudinal amplitudes, $dA_L$ and
$dA_T$, which we still need to integrate over $\beta$, ${\bf k_t}^2$ and
${\bf l_t}^2$. We use bold-faced letters to denote euclidian vectors in the
transverse plane, hence $k_t^2=-{\bf k_t}^2$, etc.

We cannot perform the ${\bf k_t}^2$ integral, as it depends on the proton
form-factor. For the ${\bf l_t}^2$ integral, we change variable to $l^2$
as this is the variable entering the vertex function.
\newcommand{\prop}{{\cal P}}
\beq \alt^2=(1-\beta^2)m_V^2-\mu_q^2+2(\beta-1)\prop\label{lt2}\eeq
As we are going to discuss the real part and the imaginary part of the
expression, we express the pole explicitly:
\beq
\prop=2l^2+{m_V^2-\mu_q^2\over 2}+i\epsilon
\eeq

Defining :\\
\begin{center}
$L^2=(1- \beta^2)(Q^2+m_V^2)-2(1-\beta)\prop$\\
$M^2=2(1- \beta^2)m_V^2-2(1-\beta)\prop$\\
$K^4= (L^2 +\akt^2)^2-4{\bf l}_t^2 \akt^2$\\
\end{center}
we obtain after angular integration:
\beqn dA_L&=& {-(1 -\beta^2)\ Q\ M^2 (K^2 -L^2)
\over (2\pi)^3\ K^2\ L^2 \akt^4 m_V\ \prop}\ dl^2\ d\beta\
d\akt^2\\
dA_T&=& {-1 \over (2\pi)^3\ K^2 L^2 \akt^4\ 2\prop}\ dl^2\ d\beta\
d\akt^2 \nonumber\\
&&\{ (K^2-L^2)[(1+\beta^2)(L^2 -2 \alt^2) -4 \mu_q^2]-\akt^2 L^2
(1+\beta^2)\ \}\nonumber\\
\end{eqnarray}

The positive energy condition $(v-l).v>{m_V \mu_q\over 4}$ implies that
\beq l^2<{(m_V-\mu_q)^2\over 4}\label{l2kin}\eeq

These expressions have the following remarkable properties:\\
i) The small-$\akt$ behaviour of both amplitudes is precisely
$1/\akt^2$. This is essential as it means that there is no IR singularity:
the remaining divergence is handled by the proton form factor. We get:
\beqn
dA_T(\akt=0)&=&{-2[(1+\beta^2) \alt^2 (\alt^2-L^2)+\mu_q^2 (2\alt^2
-L^2)]\over (2\pi)^3\ L^6 \akt^2 \prop}\ dl^2\ d\beta\ d\akt^2 \\
dA_L(\akt=0)&=&{- (1 - \beta^2)Q M^2 (L^2 -2\alt^2)\over (2\pi)^3\ L^6
\akt^2 m_V \prop}\ dl^2\ d\beta\ d\akt^2
\eeqn

\noindent ii) One recovers the ratio $ A_L/ A_T$ of our previous work, in the
limit of zero Fermi momentum:
\beqn
dA_T(\beta=0,\alt=0) &=&{2 \mu_q^2dl^2 d\beta d\akt^2 \over (2\pi)^3 (Q^2+
\mu_q^2 +\akt^2)(Q^2+
\mu_q^2) \akt^2 \prop} \\
dA_L(\beta=0,\alt=0) &=& {-\ Q\ (m_V^2 +\mu_q^2)dl^2 d\beta d\akt^2 \over
(2\pi)^3 (Q^2+
\mu_q^2 +\akt^2)(Q^2+ \mu_q^2) m_V \akt^2 \prop}
\eeqn

For $\mu_q^2=m_V^2$ and $Q>> m_V $, we recover $|{dA_L \over dA_T}|={Q \over
m_V}$.\\
iii) There is no divergence at the edge of the $\beta$ integrals.
 The
longitudinal cross section is zero, and the transverse cross section
goes to a constant, {\it independent of $Q^2$}.
\beqn
dA_T(\beta=-1)&=&{(K^2-L^2+\akt^2)\over (2\pi)^3\ K^2 \akt^4\ \prop
}\ dl^2\ d\beta\ d\akt^2\\\
dA_T(\beta=+1)&=&{ dl^2\ d\beta\ d\akt^2
\over (2\pi)^3\ \akt^2\ \prop}\times
\left[{1 \over K^2}+{(m_V^2-2 Q^2 -4 l^2 +\mu^2) \over \akt^2 (m_V^2+2 Q^2 -4 l^2
+\mu^2) }\right]
\label{betaone}\\
dA_L(\beta=\pm 1)&=&0
\eeqn

We can now perform the $\beta$ integration.
Taking into account condition (\ref{l2kin}), we write $l^2=(m_V-\mu_q)^2/4-
\lambda^2$. The positivity of $\alt^2$ from (\ref{lt2})
constrains $\beta$ to be contained between the bounds:
\beq \beta_\pm={\pm 2\sqrt{\lambda^2 + \mu_q m_V}\lambda - 2\lambda^2 - \mu_q
m_V+ m_V^2\over m_V^2}\label{betabounds}\eeq
The lower bound of integration is always comfortably below 1 for nonzero
$\mu_q$ and finite $\lambda$. The lower bound becomes smaller than -1, and is
then not realised, for
$\lambda>\sqrt{ m_V^2-m_q m_V}$.

However, the amplitude has a pole in $l^2$, or equivalently in $\lambda^2$, at
\beq \prop=0\Rightarrow l^2={\mu_q^2-m_V^2 \over 4}\label{onshell}\eeq
hence we obtain a contribution from the principal part integration, and a
contribution from the discontinuity. The latter corresponds to a direct
extension of our previous model: both quarks
constituting the meson are on-shell, and, after integrating over $\beta$, we
obtain the following for
the contribution to the amplitude at high $Q^2$:
\beqn
dA_{L(disc)}&\approx& {-8 m_V \beta_d \over \akt^2 Q^3}\ {dl^2 d\akt^2 \over (2
\pi)^3}\\
dA_{T(disc)}&\approx& {-4 m_V^2 \beta_d +2(\mu_q^2+2m_V^2) \log({1 + \beta_d
\over 1-\beta_d})\over \akt^2 Q^4}\ {dl^2 d\akt^2 \over (2 \pi)^3}
\eeqn
The bounds on $\beta$ corresponding to (\ref{betabounds}) are
$\beta_d^2={m_V^2-\mu_q^2 \over m_V^2}$, far away from $\beta=\pm 1$. We see
that the ratio $ dA_L/ dA_T$ is still linear, but depends on the quark
mass chosen. In the limit
$\beta_d\rightarrow 0$, we of course recover our previous results for the
ratio, and that the ratio $Q/m_V$ gets multiplied by a constant, which depends
on the quark mass. Unfortunately, this constant is always between 0.5 and 1, as
shown in Fig.~2 for reasonable values of the (constituent) quark mass.  Hence
the discontinuity alone cannot solve the problem of the ratio
$\sigma_L/\sigma_T$.
\begin{figure}[here]
\label{onshellratio}
\centerline{
\hskip -2cm
\psfig{figure=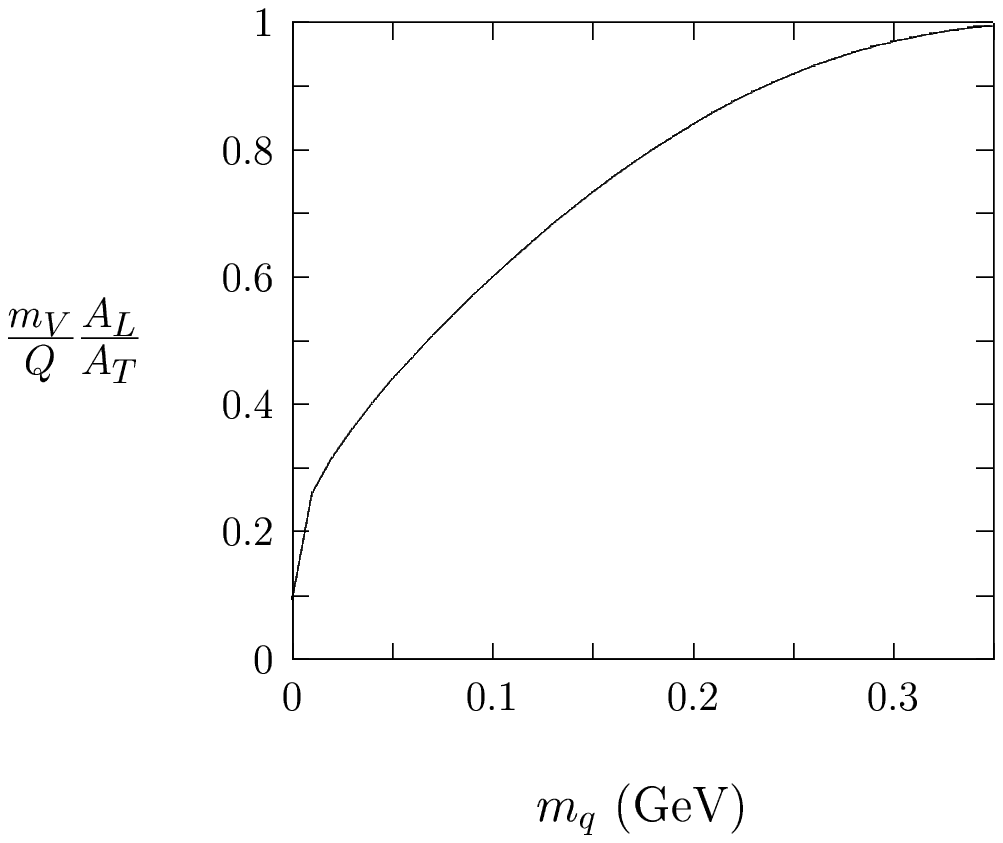,width=8cm}}
\begin{quote}
{\small Figure~2: The reduction factor in the ratio $\sigma_L/\sigma_T$ as a
function of the quark mass
}\end{quote}
\end{figure}

Note that it is possible to extend somewhat the previous expressions. Although
the expressions include Fermi motion, they really single out a value of $l^2$
which corresponds to the on-shell condition, and the remaining $\beta$ integral
can be
reduced to an angular integral in the meson center-of mass frame. One can
make another model, where the quark mass is a priori not fixed, but a function
of $l^2$, chosen so that the on-shell condition (\ref{onshell}) is satisfied.
The remaining integration involves then the vertex function
$\Phi\left(l\right)$, and
it cannot be done analytically. We have tried this route numerically, and find
that the ratio remains linear, and can be somewhat reduced, by about a factor
2.
This is not enough to reproduce the measured values, and is reminiscent of the
situation encountered using a light-cone wavefunction formalism.

At this point, we still have not included the contribution of the principal
part integration. This corresponds to letting one of the quarks off-shell,
and there is no reason to neglect it. In fact, in the case of DIS, the
kinematics is such that this is the only contribution. Its structure is
somewhat different from that of the discontinuity. Indeed, the value of $l^2$
is not fixed anymore,
hence the integration over $\beta$ has only the bounds (\ref{betabounds}).
In particular, once $l^2$ becomes large and negative, the value $\beta=-1$
can be realised. As we have mentioned in Eqs.~(\ref{betaone}), the transverse
part retains a finite value there, whereas the longitudinal part goes to zero.
Hence there is a narrow peak which gives a contribution only in the transverse
case. If we include only the large $\lambda$ contribution (the other parts turn
out to be negligible), one obtains:
\beqn
 dA_{L(PP)}&=&{dl^2\ d\akt^2 \over (2\pi)^3\ \akt^2\ m_V Q^3
}
\times\left\{{2m_V^2 (1+\beta_+)\over \prop} \right.\nonumber\\&+&
\left.\log\left[{(4 \prop-\akt^2)\ \prop
\over (1+\beta_+)^2 Q^4}\right]
-4{\prop \over \akt^2} \log\left[{ 4 \prop-\akt^2
\over 2 \prop} \right]\right\}\label{PP}\\
dA_{T(PP)}&=&{-2\ \log\left({4\prop-\akt^2 \over 4\prop}\right)\over \akt^4\
Q^2}{dl^2\ d\akt^2 \over (2\pi)^3}
\label{PPT}\end{eqnarray}

We can see that the real parts behave like $ dA_L \propto {1 \over Q^3}$,
and $ dA_T \propto {1 \over Q^2}$ at high $Q^2$, plus logarithmic
corrections, whereas the imaginary parts still behave like $ dA_L \propto {1
\over Q^3}$, and $ dA_T \propto {1 \over Q^4}$. Because this effect is
present only at large $\lambda$, it
is suppressed by the fall-off of the vertex function, and sets in only at
relatively large $Q^2$. The leading behaviour of the principal parts integrals
comes from quark off-shellnesses bigger than the constituent quark mass.
Whereas in the massless case \cite{Collins} an increase in the cross section
can only come from extremely small off-shellnesses, we find that the dominant
region in the massive case is shifted by the quark mass: the principal part
integral cancels as long as one is very close to the pole, and the kinematics
is such that this cancellation is not present anymore once the off-shellness is
of the order of the quark mass.

Hence the source of the plateau observed at HERA is the interplay between the
real part and the imaginary part of the cross section. Our model in fact
predicts that asymptotically the transverse and the longitudinal cross sections
first become equal, and that ultimately the process is dominated by the
transverse cross section. This prediction is however driven by the details of
the quark propagators, which could be modified by confinement effects.

\section{Results}
We shall now give our numerical results, which come from a numerical study of
the full integrands at all $t$ values. Before doing so, though, we shall need
to define the vertex function, and to normalise it.

\subsection{Vertex function and normalisation}
The function $\Phi$ is unknown, and only its general analytical properties are
well established \cite{LandHe}. In general, this function can depend on the
scalar products $V.V$, $V.l$ and $l.l$. However, as in our case
one of the quarks is on-shell, and as the meson is on-shell,
only one of these scalar products is free. As in the case of the imaginary
part, our results reduce to those of a wavefunction formalism, we shall
assume that the vertex function is similar to the wavefunction of an s state.
In the case of the $\rho$, the J/$\psi$ and the $\phi$, we take the form:
\beq
\Phi\left(l\right)=N\ e^{-{{\bf L}^2 \over 2 p_f^2}}
\label{L}
\eeq
where ${\bf L}^2$ is the quark 3-momentum in the meson rest frame, equal to
${\bf L}^2=({l.V \over m_V})^2-l.l$, where  the Fermi momentum $p_F$ is
0.3 GeV in the $\rho$ and $\phi$ cases, and 0.6 GeV in the J/$\psi$ case
\cite{fermimom}, and where the quark masses are taken to be $m_{u,d}=0.3$ GeV,
$m_s=0.45$ GeV and $m_c=1.5$ GeV.

We can now fix the normalisation constant $N$ by requiring that the decay rate
$\Gamma\rightarrow e^+e^-$ be reproduced through the process depicted in
Fig.~3.

\begin{figure}[here]
\label{Normalisation}
\centerline{
\hskip -2cm
\psfig{figure=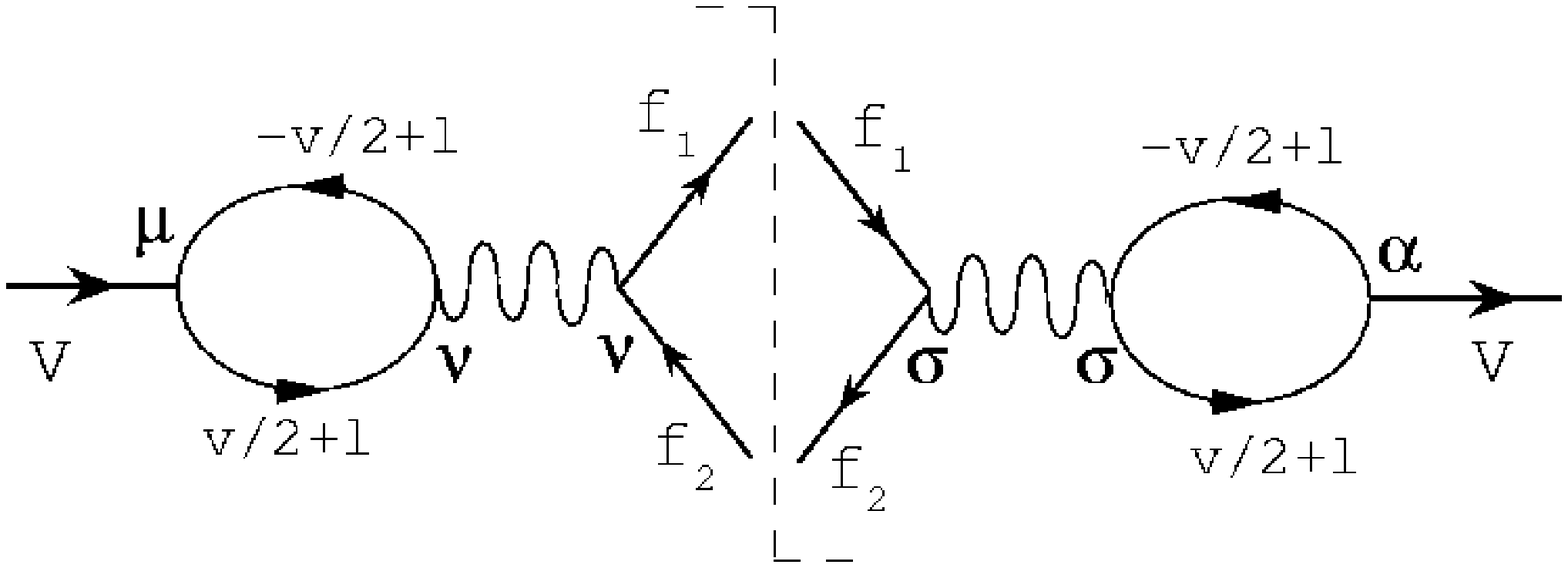,bbllx=4.5cm,bblly=14cm,bburx=15cm,bbury=19cm,width=5cm}}
\begin{quote}
\vskip 1in
{\small Figure~3:  Diagram renormalising the vertex
}\end{quote}
\end{figure}

The total decay rate, $\Gamma$, for $V\rightarrow f_1 f_2$ is given by:
\beq
\Gamma={1 \over 2{m_V}}
\int {{{d^4f_1 d^4f_2\over {(2\pi)^2}}}|{\it M}|^2 \delta^{(4)}(V-f_1-f_2)
\delta({f_1}^2-{m_f}^2) \delta({f_2}^2-{m_f}^2)}.
\eeq \\
with $M$ the invariant amplitude corresponding to the diagram of Fig.~3:
\begin{eqnarray}
|{\it M}|^2&=&\int {{d^4l \over {(2\pi)^4}}\ T_1\  {g_{elm}}^2\ e_Q\ ({3 \over
\sqrt{3}})\ \Phi\left(l\right)}\nonumber\\
&\times& (T_2) \nonumber\\
&\times& \int {{d^4{l'} \over {(2\pi)^4}}\ T_3\  {g_{elm}}^2\ e_Q\ ({3 \over
\sqrt{3}})\ \Phi\left(l\right)}\nonumber\\
&\times& ({1\over V^2})^2
\end{eqnarray}\\
With
\begin{eqnarray}
T_1&=&{Tr [\gamma.e (\gamma.(-v+l)+m_q) \gamma^\nu (\gamma.(v+l)+m_q)]
 \over {[(-v+l)^2-m_q^2][(v+l)^2-m_q^2]}}\\
T_2&=&Tr [\gamma^\nu (\gamma.f_2-m_f) \gamma^\mu (\gamma.f_1-m_f)]\\
T_3&=&{Tr [\gamma^\mu (\gamma.(-v+l)+m_q) \gamma.e (\gamma.(v+l)+m_q)]
\over {[(-v+l)^2-m_q^2][(v+l)^2-m_q^2]}}
\end{eqnarray}\\

Putting the final leptons on-shell, neglecting their mass, working in the
center-of-mass frame of the meson, where $l^2=l_0^2-{\bf L}^2$, and performing
the angular integrals, leads to:
\begin{eqnarray}
\Gamma&=&{1 \over {2m_V}}\ {1 \over {(2\pi)^2}}\ {{\pi V^2} \over {24}}\ {9
\over 3}\ {g_{elm}}^4\ {e_Q}^2\ ({1 \over {V^2}})^2\ ({128 \over 3})^2\
\nonumber\\
&\times&\left[\int{{{{\bf L}^2 dl_0 d{\bf L}} \over (2\pi)^3}\
\frac{[3(4m_q^2+m_V^2)+4{\bf L}^2-12l_0^2]\ \Phi\left(l\right)}{(4l_0^2+4l_0
m_V -4{\bf L}^2-4m_q^2+m_V^2+i\epsilon )}}\right.\nonumber\\
&\times& \left.{1\over (4l_0^2 -4l_0 m_V-4{\bf L}^2-4m_q^2+m_V^2+i\epsilon
)}\right]^2
\end{eqnarray}

The integration over $l_0$ can be carried out by residues. We then get:
\begin{eqnarray}
\Gamma&=&{1 \over {2m_V}}\ {2\pi \over 3}\ {\alpha_{elm}}^2\ {e_Q}^2\ {9 \over
3 m_V^2}\ \left[{8 \over {3(2\pi)^3}}\ (2i\pi)^2\right]^2\nonumber\\
&\times& \left[\int {\frac {{\bf L}^2 d{\bf L}}{\sqrt{{\bf L}^2+m_q^2}}\
\frac{(2{\bf L}^2+3m_q^2)\ \Phi\left(l\right)}{({\bf L}^2+m_q^2-{m_V^2 \over
4})}}\right]^2
\end{eqnarray}

The remaining integral over ${\bf L}$ still has a pole at the quark propagator.
We must again keep both the discontinuity and the principal part of the
integral. We in fact see that the phase of the amplitude becomes pure imaginary
again
once the vertex gets properly normalised. Introducing the function
 we have to deal with the pole in the denominator for ${\bf L}=\sqrt{{m_V^2
\over 4}-m_q^2}$.\\
We introduce the notations:
\begin{eqnarray}
&&z=2{\bf |L|}\nonumber\\
&&z_d=\sqrt{m_V^2-\mu_q^2}\nonumber\\
&&f(z)=\frac {z^2}{\sqrt{z^2+\mu_q^2}}\ \frac{(2z^2+3\mu_q^2)\
\Phi(z^2/4)}{(z-z_d)}
\end{eqnarray}
with $\Phi(z^2/4)=N\ e^{{-z^2 \over 8 p_f^2}}$
then the integration over {\bf L} gives:
\beq
\Gamma={4\alpha_{elm}^2\ {e_Q}^2\over 9m_V^3\pi}
\left[P\int_{0}^{\infty} {f(z)dz\over z-z_d}+ (i\pi)\ f(z_d)\right]^2
\eeq
It is worth pointing out that if we neglect the momentum $l$ in the quarks
loop, e.g. for $m_q={m_V \over 2}$, we recover the formula \cite{Horgan} for
the $\rho$ meson:
\beq
\Gamma={f_m^2 \over m_V}\ {8\pi \over 3} \alpha_{elm}^2\ e_Q^2
\eeq 
We give in the Table the values of the decay rates \cite{PDG} which we have
fitted to, and the corresponding values of the normalisation $N$.\\
\begin{quote}
\begin{center}
\begin{tabular}{|c|c|c|c|c|}\hline
Meson &$\Gamma(V\rightarrow e^+e^-$) (keV)&$|N|^2$&$p_F (GeV)$ & $m_q$ (GeV)\\ \hline
$\rho$&6.77&61.71&0.3&0.3\\
$\phi$&1.37&72.01&0.3& 0.45\\
$J/\psi$&5.26&44.91&0.6& 1.5\\\hline
\end{tabular}
\end{center}
{\small Meson decay rates, vertex normalisation constants, Fermi
momenta and quark masses.}
\end{quote}
\subsection{$Q^2$ and mass dependence of the cross section}
The $Q^2$ behaviour of the total cross section, as well as the mass dependence,
hardly get affected by the addition of Fermi momentum. This is because at low
$Q^2$, the transverse amplitude is not modified by the addition of the real
part, whereas at high $Q^2$ the cross section is still dominated by the
longitudinal part at present values of $Q^2$. We show in Fig.~4.a the result of
our model together with data from HERA. We have
not corrected either for an eventual mass or $Q^2$ dependence of the Regge
factor.
Also, as our model does not include a prediction of the energy dependence, we
have compared with 1994 data. New data are more precise, and allow one to
observe that energy dependence directly from HERA data. The figure (plotted
for data at $w^2\approx 100$ GeV$^2$) shows
however that the general features of the data are well reproduced:  we see that
a constant Regge factor (equal to 5.1) reproduces the data well for all mesons
at
HERA. This can be checked by considering figure 4.b which
shows the ratio of cross sections, which goes well through the absolute
predictions of our model. As in our previous model \cite{CR1} we see that there
seems to be a deficit of $\phi$ mesons w.r.t. $\rho$ mesons at NMC, which we
are unable to account for. Comparison with lower-energy data may lead to an
estimate
of $\alpha_S$ as well as of the intercept. These estimates are identical to
those of \cite{CR1} although the value of $\alpha_S$ should be multiplied by a
factor $\sqrt{3}$ which we overlooked in our previous work \cite{Diehl}.
It is again possible to go through the photoproduction point, although
admittedly our model should not work for such low values of $Q^2$.
\begin{figure}
\vglue 1cm
\hglue 4.5cm (a)\hglue 4.5cm (b)\\
\vglue -1.5cm
\centerline{
\hbox{\psfig{figure=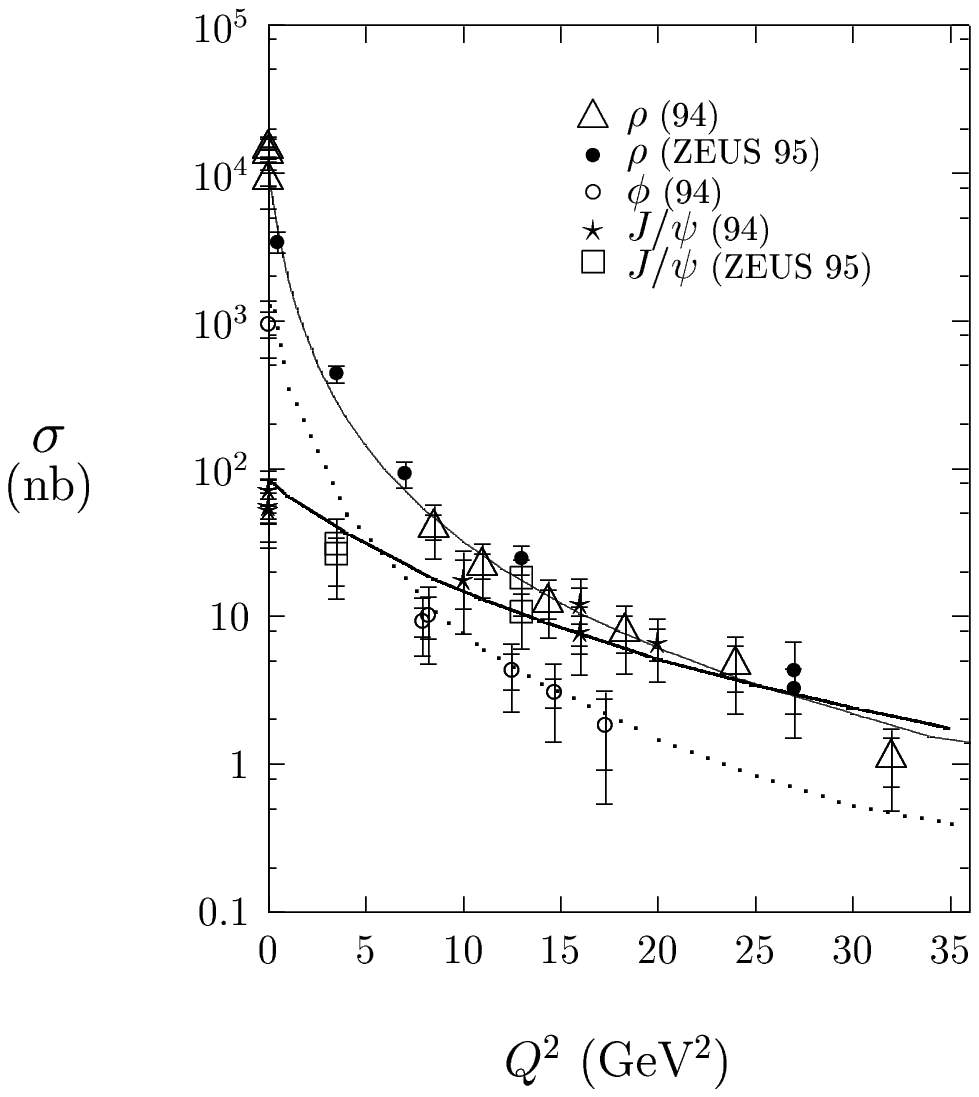,height=5.5cm}}\ \ \
\hbox{\psfig{figure=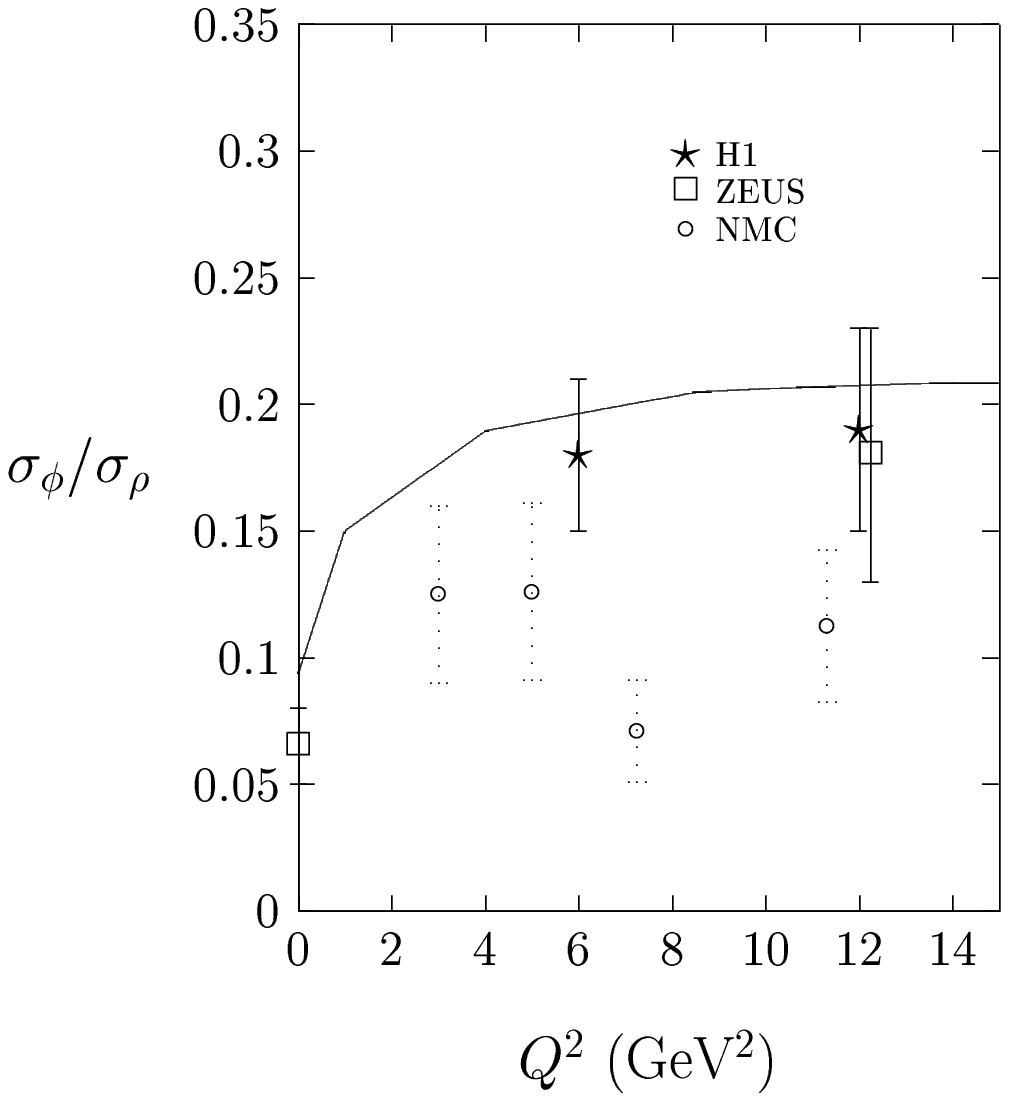,height=5.5cm}}}
\begin{quote}
{\small Figure~4: (a) Cross sections for $\rho$ as functions of $Q^2$, compared
with data from ZEUS~\cite{Zeusdat} and H1~\cite{H1dat}
at $<w>\approx$100~GeV, (b) Ratio of cross sections as functions of $Q^2$
at \hbox{$<w>\approx$100~GeV}, compared with data from H1
and Zeus \cite{H1dat,Zeusdat}, and from NMC \cite{NMC}.
}\end{quote}
\end{figure}

We have explained in great detail at $t=0$ that ratio of the longitudinal and
transverse amplitudes has a plateau at large $Q^2$. It remains to be seen
whether this feature is maintained for the $t$-integrated cross section. We
show in Figure~5.a that this is indeed the case. Furthermore, as demonstrated
in
Fig.~5.b, for the values of Fermi momentum and for the quark masses shown in
the Table, one falls right on top of the data. There is of course some
uncertainty linked to these values, but a change of 50\% in either leads
to a change of about 20\% in the ratio $\sigma_L/\sigma_T$.
\begin{figure}
\vglue 1cm
\hglue 4.5cm (a)\hglue 7.5cm (b)\\
\vglue -1.3cm
\centerline{
\hbox{
\psfig{file=P,bbllx=3.8cm,bblly=13.6cm,bburx=17.2cm,bbury=25.1cm,height=5.5cm}
}\ \ \
\hbox{
\psfig{file=R,height=5.5cm}
}
}
\begin{quote}
{\small Figure~5:  (a) Scaled behaviour of the real and imaginary parts of the
transverse and longitudinal amplitudes, (b) Ratio of the longitudinal and
transverse parts of the $\rho$ cross section as functions of $Q^2$}
\end{quote}
\end{figure}

\subsection{Photoproduction, Regge factor and $t$ dependence of the cross
section}
As we briefly mentioned, our model works in the deep non-perturbative region of
photoproduction. It may be worth mentioning at this point that even at large
values of $Q^2$ the gluon off-shellnesses are not large, and as we have seen
quark off-shellnesses must be near $m_q$ in the real part. Hence the
situation is not dramatically different in the case of photoproduction, but
nevertheless, it still comes as a surprise that the model applies in this
region. This may hint at a lower-$Q^2$ generalisation of the factorisation
theorem.

We show in Fig.~6.a the $\rho$ photoproduction cross section $d\sigma/dt$ and
compare it with our model. We see that we obtain excellent agreement, although
our curve is not an exponential. In Fig.~6.b, we show the $Q^2$ dependence of
the $b$ slopes from our model: the experimental points correspond to a fit
to $Ne^{bt}$, and our curve corresponds to 1/$<t>$, which would be the same
were the curve an exponential. We see that despite the curvature of our
curves, the agreement is quite good.
\begin{figure}
\vglue 1cm
\hglue 4.0cm (a)\hglue 5.5cm (b)\\
\vglue -1.5cm
\centerline{
\psfig{figure=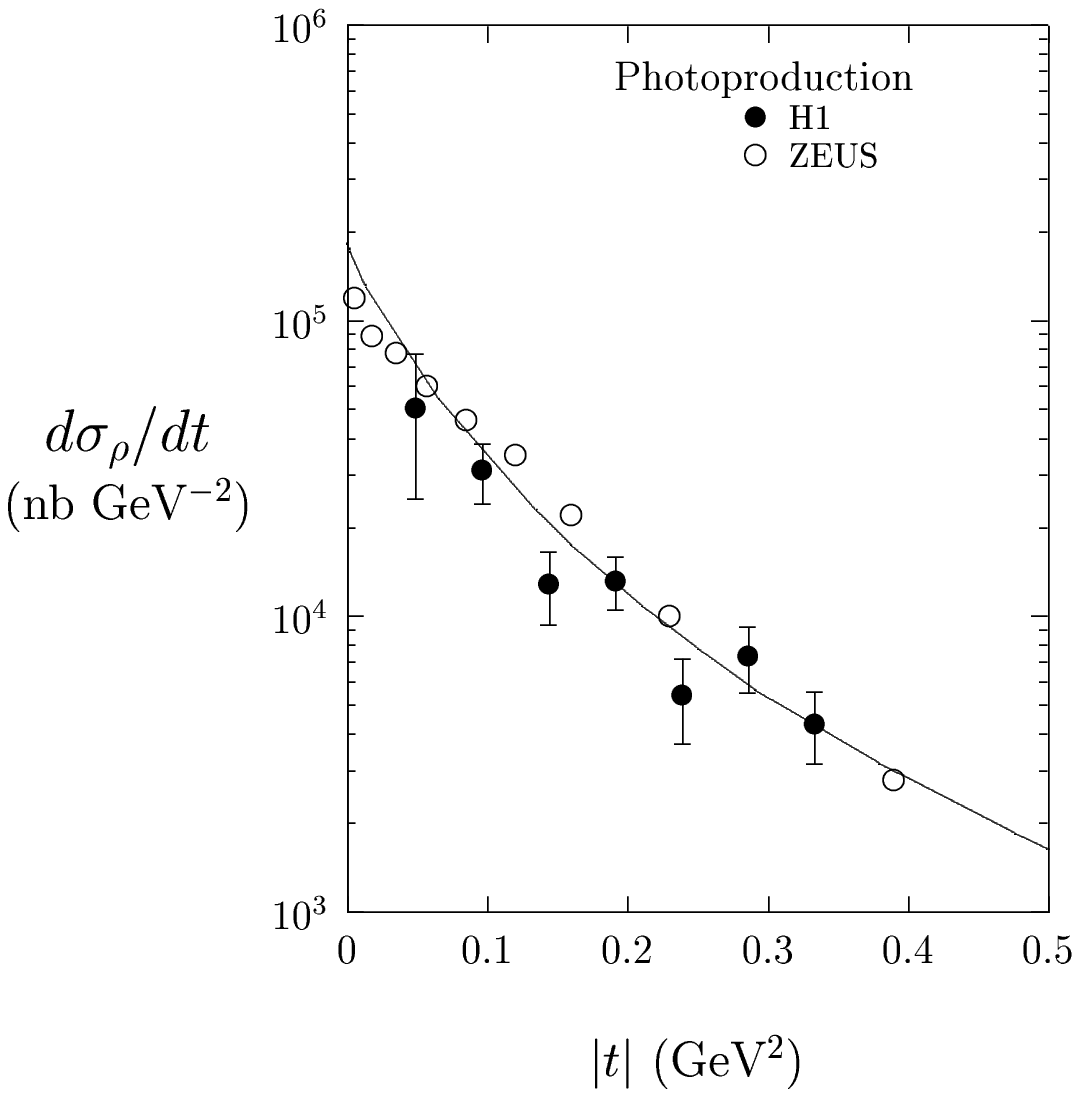,height=5.5cm}\ \ \ \ \
\psfig{figure=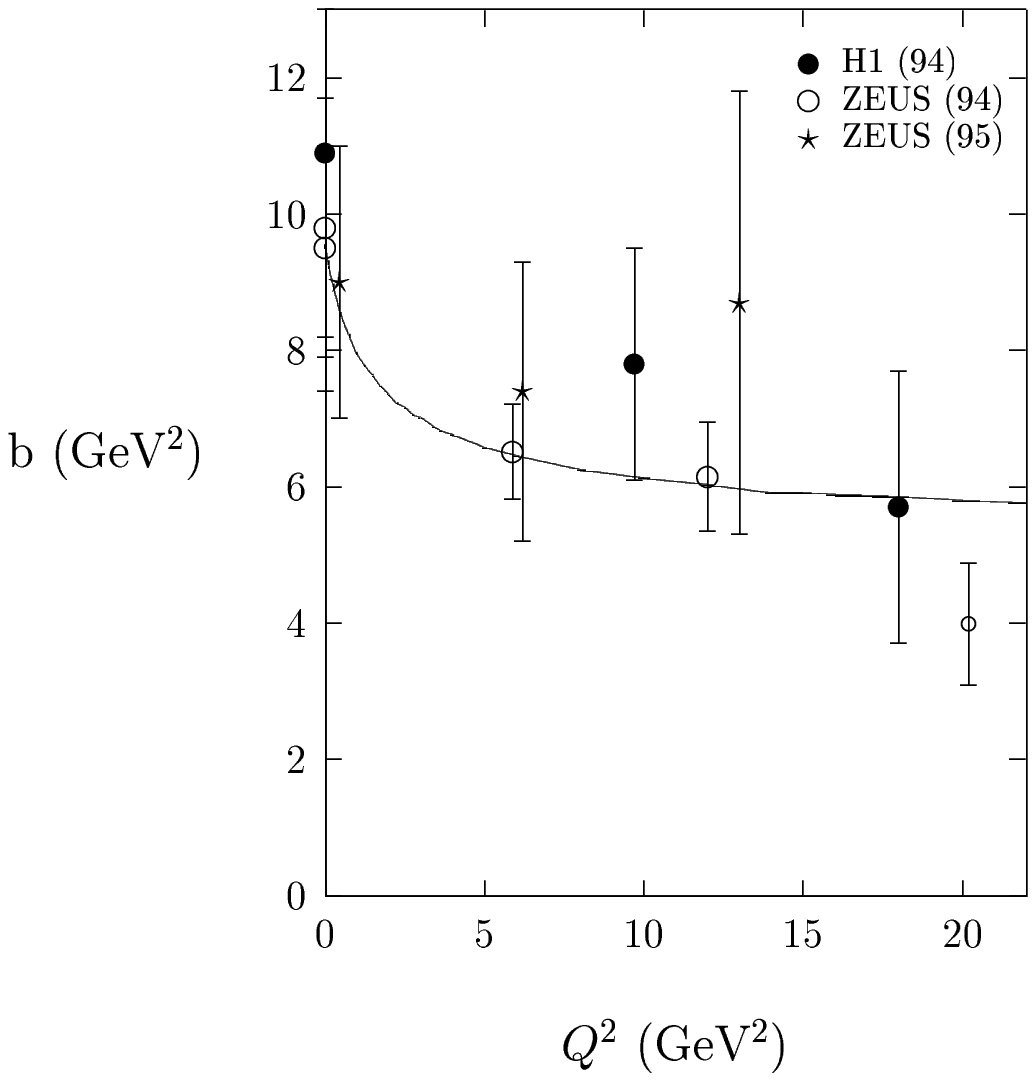,height=5.5cm}}
\begin{quote}
{\small Figure~6: (a) Differential cross section $d\sigma/dt$ for elastic
$\rho^0$ photoproduction, compared with HERA data \cite{Zeusdat,H1dat},
(b) $t$-slope of the photoproduction differential elastic cross sections,
compared with HERA data \cite{Zeusdat,H1dat}.
}\end{quote}
\end{figure}
The $t$-dependence of other mesonic cross sections is predicted in
Fig.~7. Comparison with preliminary ``public'' data from ZEUS
\cite{crittenden} would indicate good agreement. We trust that
the forthcoming ZEUS published analysis will confirm this result.
\begin{figure}
\vglue 1cm
\hglue 4.0cm (a)\hglue 5.5cm (b)\\
\vglue -1.5cm
\centerline{
\psfig{figure=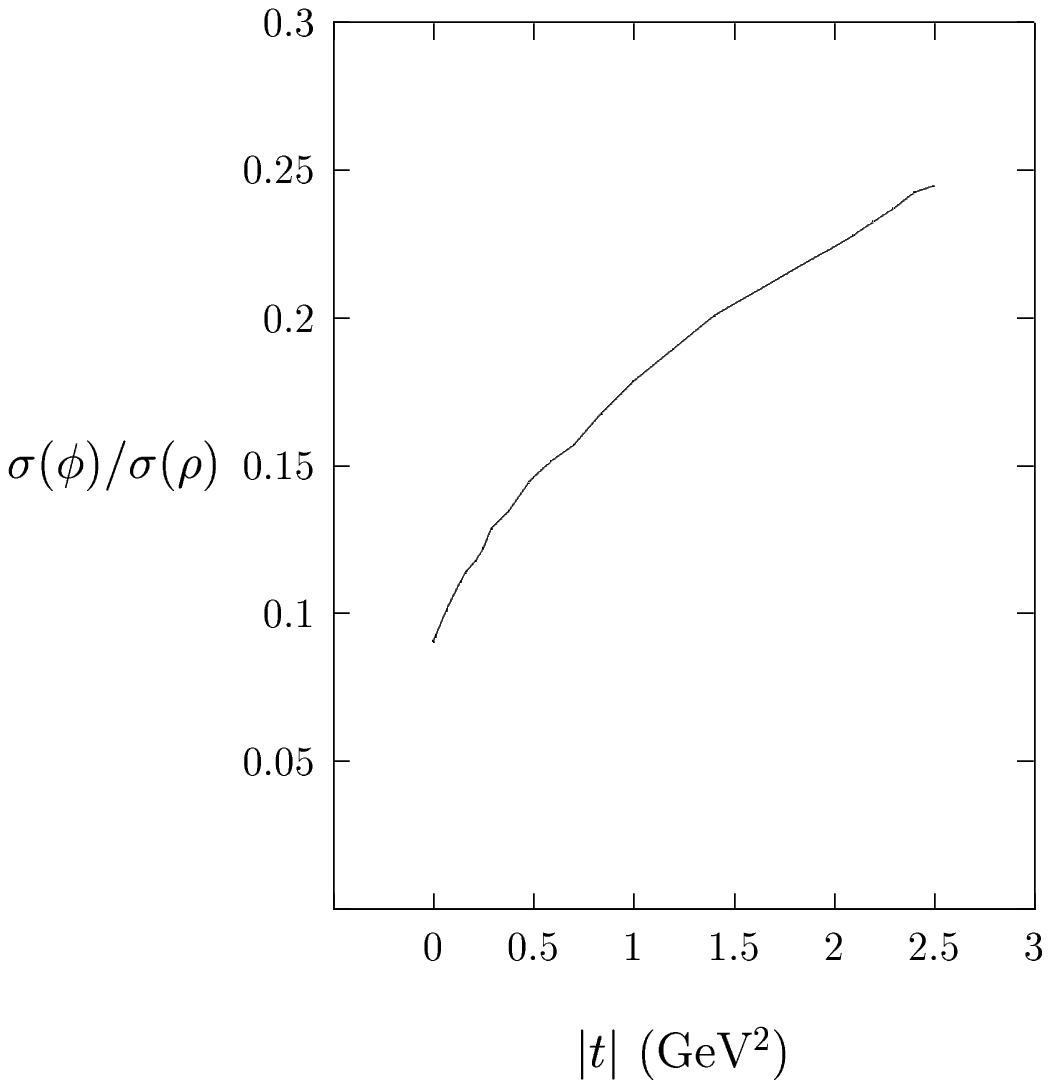,height=5.5cm}\ \ \ \ \
\psfig{figure=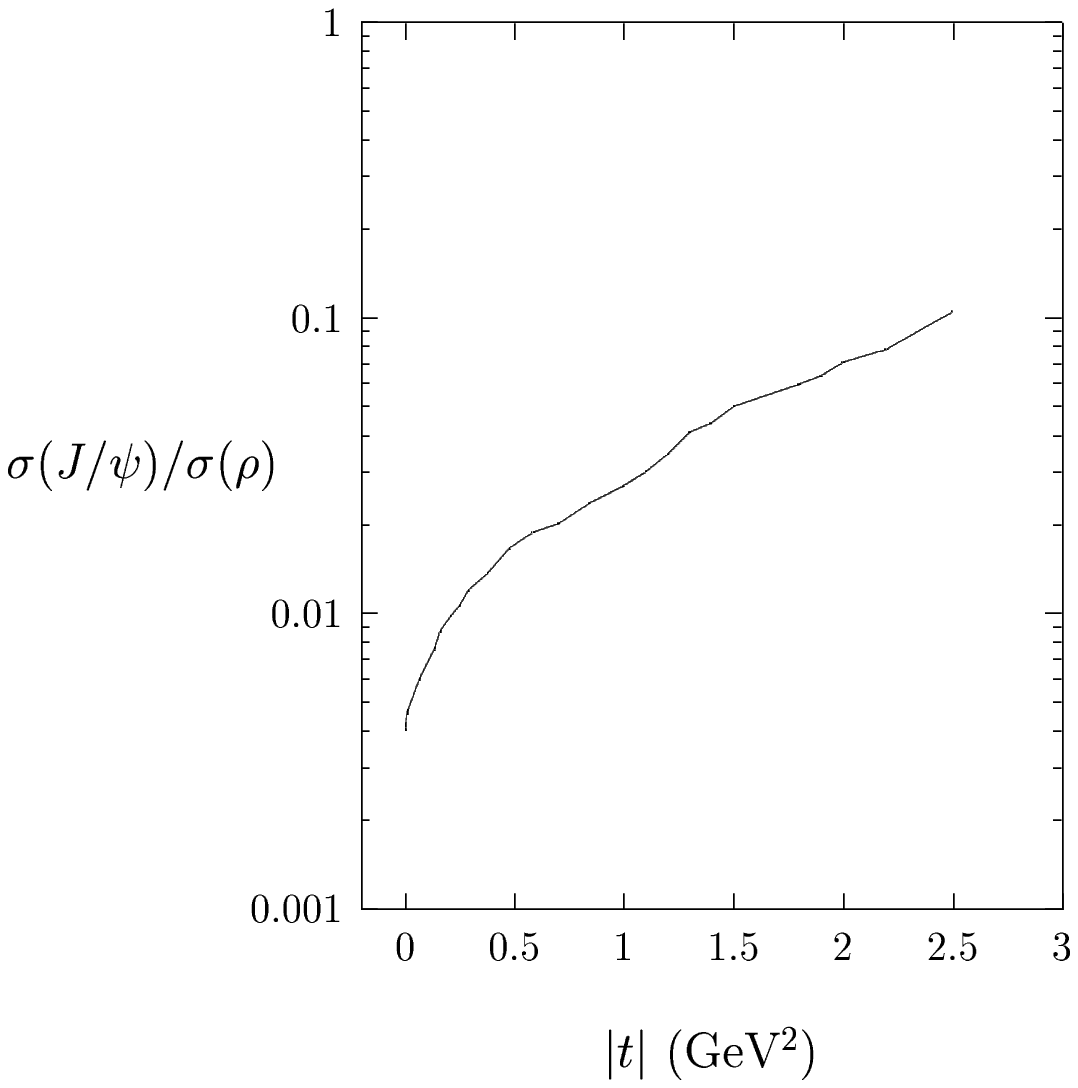,height=5.5cm}}
\begin{quote}
{\small Figure~7: Ratio of the differential photoproduction cross sections
for (a) $\phi$ and $\rho$ mesons and (b) for $J/\psi$ and $\rho$ mesons.
}\end{quote}
\end{figure}
Hence, if we believe this model is applicable to photoproduction, we see there
is little room either for a BFKL large $t$ enhancement or for a pomeron slope.

\subsection{Predictions for $\rho'$ and $\Psi'$}
Given the success of our model in reproducing the lowest mass vector mesons, we
can easily extend it to study the 2s excited states. The first ingredient
is the vertex function, which we take as:
\begin{equation}
\Phi_{2s}=N\sqrt{2/3}\left(-{3\over 2}+{{\bf L}^2\over p_F^2}\right)
\exp\left(-{{\bf L}^2\over 2p_F^2}\right)
\eeq
with $\bf L$ as defined below Eq.~(\ref{L}), and the values of $p_F$ and $m_q$
the same as in the 1s case. The normalisation constant $N$ should again
be determined from the leptonic decay of the vector meson. The latter
is available only in the $\psi'$ case \cite{PDG}, hence we can make predictions
only
in this case: $\gamma(\psi'\rightarrow e^+e^-)=5.26$ keV leads to $N=35.36$.
The ratio of the $\psi'$ elastic production to that of the $J/\psi$, at low
$Q^2<0.01$ GeV$^2$ is calculated to be 0.165, in excellent agreement with the
H1 measurement \cite{prime}, as can be seen from Fig.~8.a.
\begin{figure}
\vglue 1cm
\hglue 3.6cm (a)\hglue 5.5cm (b)\\
\vglue -1.5cm
\centerline{
\psfig{figure=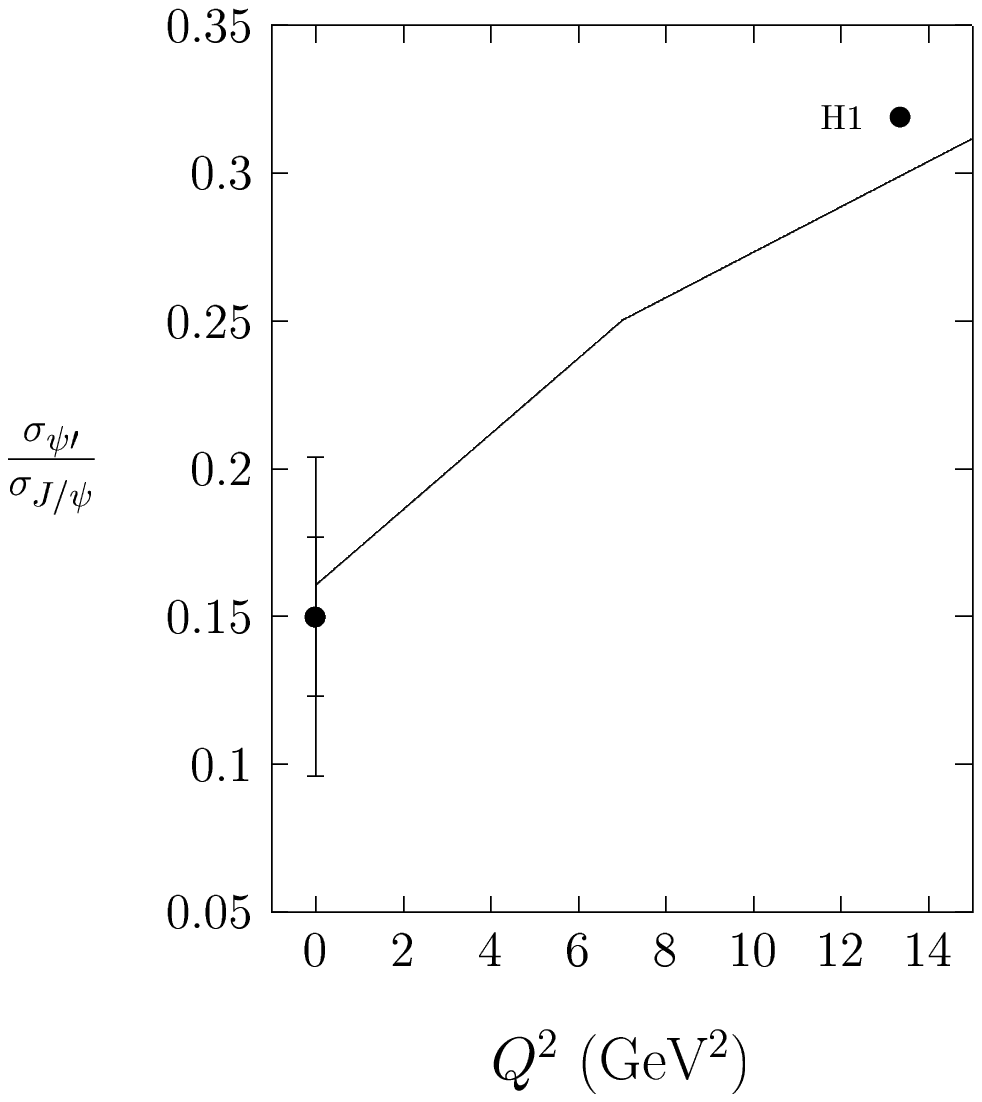,height=5.5cm}\\
\psfig{figure=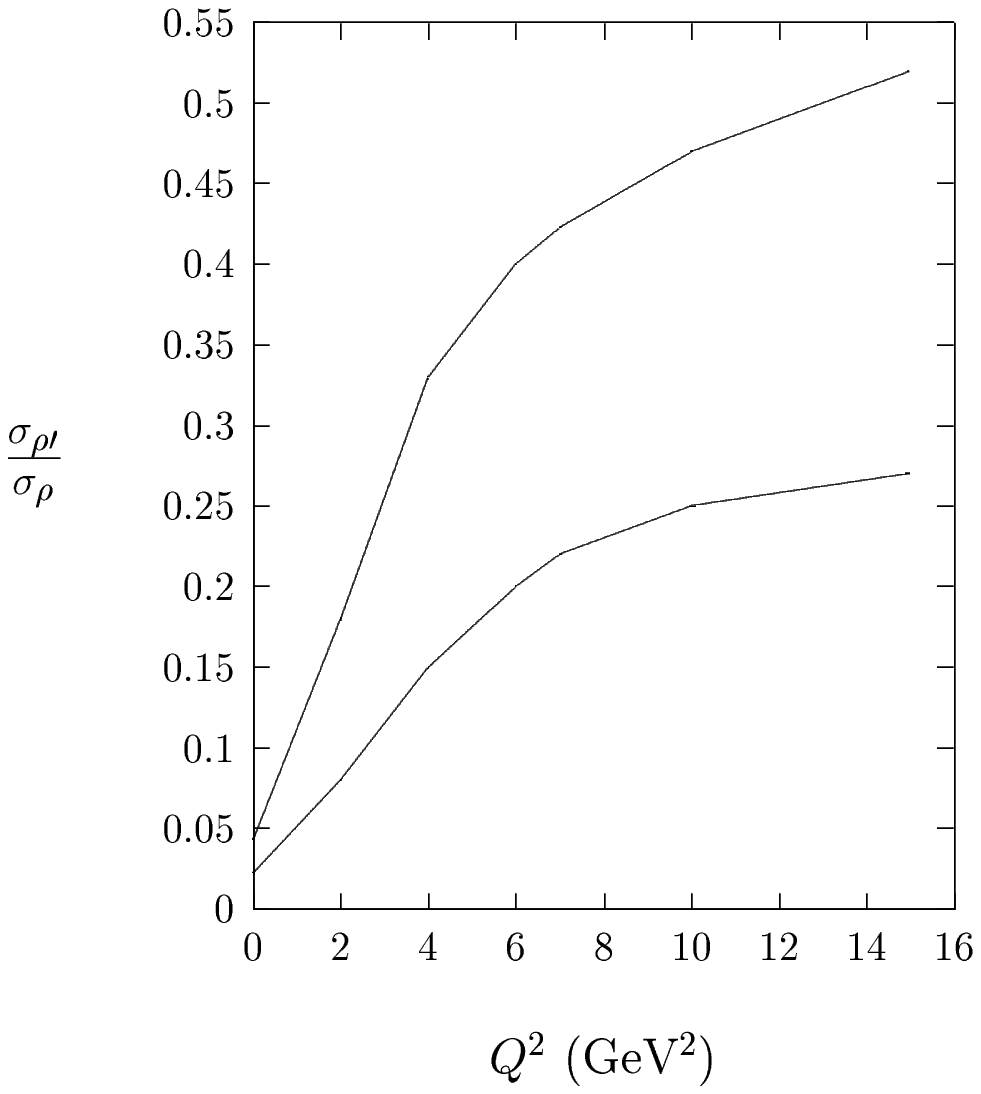,height=5.5cm}}
\begin{quote}
{\small Figure~8: (a) Prediction for the ratio of the $\psi'$
production cross section to that of the $J/\psi$, compared with
H1 data \cite{prime} (b) prediction for the $\rho'$ cross section
based on the consideration of H1 preliminary ``public'' data \cite{rhoprime},
leading to the prediction of the leptonic width (\ref{lepto}).
}\end{quote}
\end{figure}

In the $\rho'$ case,
we can calculate $N$ from the elastic production and make predictions for the
leptonic width: from the preliminary ``public'' results of the H1 collaboration
\cite{rhoprime}, which we were unfortunately asked not to 
quote \cite{Dainton}, we obtain
\beq \Gamma(\rho'\rightarrow e^+e^-)=1.1\pm 0.34 \ \rm keV\label{lepto}\eeq
The resulting predictions are shown in Fig.~8.b. We see that although
the normalisation of this figure is not predicted, the ratio is predicted
to increase sharply with $Q^2$ 
and hence an eventual final measurement at HERA should find a substantially
higher cross section than the one at  fixed
target experiments.

\section{Conclusion}
We have presented here a model which reproduces all the features (except the
$w^2$ dependence) of
elastic vector meson production as observed at HERA. Although the
model is built to work at large values of $Q^2$, it extends to the
nonperturbative photoproduction region. The main result is that it is possible
to
reproduce the ratio $\sigma_L/\sigma_T$. The plateau observed experimentally
comes from the interplay between the on-shell and the off-shell quark
contributions, which have different asymptotic behaviours. Hence it is
essential
to allow quarks to be virtual, as in the standard DIS case.

It is worth pointing out that despite its successes, our model has not been
tuned to reproduce the data: the values of the Fermi momentum, the form of the
wavefunction and the quark masses are given in the Table, and are only
reasonable guesses. Similarly, the proton form factor could be modified, as it
is known only in the IR limit. We find remarkable that educated guesses lead to
such a good agreement with data.

Hence it is essential to allow quarks to be off-shell not only to reproduce the
behaviour of the transverse part, but also because the relation to structure
functions is only possible then. Indeed, the $\log Q^2$ terms of Eq.~(\ref{PP})
are the same as those of the upper quark loop in DIS. The use of wavefunctions
can miss such contributions, as they come from a part of the amplitude which is
usually embedded in the wavefunction.

{}~\\~\\~\\
{\noindent \Large \bf Note \hfil}\\ ~\\
A previous version of this paper showed preliminary H1 and ZEUS data scanned 
from ``public'' documents available from the web and comparing them with our model. 
Although we found beautiful agreement, 
the H1 collaboration demanded that we remove these
from the published version, and the ZEUS collaboration indicated
that they as well do not want preliminary results included for comparison 
in theoretical papers. 
We have thus removed these data, and deeply regret to have taken the 
scientific liberty of explaining experimental results publicly
available before they are published.
We want to indicate that that policy should be spelled out explicitly
in the papers and in the webpages available, as comparing results
with  
(even unpublished) data is -and should be-  a common trend among theorists.
{}~\\~\\~\\

{\noindent \Large \bf Acknowledgments \hfil}\\ ~\\
We thank Markus Diehl for several essential discussions, and for
correcting some of our formulae, Roberto Sacchi
(Zeus) for giving us
the files connected with his analyses of the data, 
J. Gravelis and
A. Donnachie for spotting a mistake in an earlier version
of this paper, Barbara Clerbaux for making us aware of the 
rules of the H1 collaboration concerning preliminary data,
and Bruno Van Den Bossche
for sharing his knowledge of meson wave functions with us.

\end{document}